\documentclass[12pt]{elsarticle}
\usepackage{amssymb}
\renewcommand\a{\alpha}
\renewcommand\b{\beta}
\renewcommand\d{\delta}
\renewcommand\k{\kappa}
\renewcommand\l{\lambda}
\renewcommand\r{\rho}

\renewcommand\t{\tau}

\renewcommand\c{\chi}

\renewcommand\o{\omega}
\newcommand\e{\epsilon}
\newcommand\g{\gamma}
\newcommand\z{\zeta}
\newcommand\m{\mu}
\newcommand\n{\nu}
\newcommand\x{\xi}
\newcommand\p{\pi}
\newcommand\h{\theta}
\newcommand\s{\sigma}
\newcommand\f{\phi}
\newcommand\w{\eta}
\newcommand\ve{\varepsilon}

\newcommand\vf{\varphi}

\renewcommand\O{\Omega}
\renewcommand\H{\Theta}
\newcommand\D{\Delta}

\newcommand\F{\Phi}

\newcommand{\eq}[1]{Eq.~(\ref{#1})}
\newcommand{\eqs}[2]{Eqs.~(\ref{#1})-(\ref{#2})}

\newcommand\lb{\left(}
\newcommand\rb{\right)}
\newcommand\ls{\left[}
\newcommand\rs{\right]}

\newcommand{\lan}{\langle}
\newcommand{\ran}{\rangle}

\newcommand\ra{\rightarrow}

\newcommand{\non}{\nonumber\\}
\newcommand\pt{\partial}

\newcommand{\ie}{\emph{i.e.}}

\newcommand{\eg}{\emph{e.g.}}

\newcommand{\diag}{{\rm{diag}}}

\newcommand{\Tr}{{\rm Tr}}

\newcommand{\im}{{\rm{Im}}}

\newcommand{\bx}{{\mathbf x}}

\newcommand{\bk}{{\mathbf k}}

\newcommand{\pperp}{{P_\perp}}
\newcommand{\ppara}{{P_\parallel}}
\newcommand{\zperp}{{\z_\perp}}
\newcommand{\zpara}{{\z_\parallel}}
\newcommand\lv{\left|}
\newcommand\rv{\right|}
\newcommand{\rh}{{\hat \r}}
\newcommand{\ah}{{\hat A}}
\newcommand{\bh}{{\hat B}}
\newcommand{\zh}{{\hat Z}}
\newcommand{\ch}{{\hat C}}
\newcommand{\gh}{{\hat G}}
\newcommand{\kh}{{\hat K}}
\renewcommand{\th}{{\hat T}}
\newcommand{\nh}{{\hat N}}

\journal{Annals of Physics (NY)}
\begin{document}

\begin{frontmatter}
\title{ Kubo formulas for relativistic fluids in strong magnetic fields
}
\author{\normalsize{Xu-Guang Huang$^{1}$, Armen Sedrakian$^1$, and Dirk H.\ Rischke$^{1,2}$}}
\address{
$^1$ Institute for Theoretical Physics, J. W.\ Goethe-Universit\"at,
D-60438 Frankfurt am Main, Germany,\\
$^2$ Frankfurt Institute for Advanced Studies, D-60438 Frankfurt am Main, Germany
}

\date{\today}

\begin{abstract}
Magnetohydrodynamics of strongly magnetized relativistic fluids is
derived in the ideal and dissipative cases, taking into account the breaking
of spatial symmetries by a quantizing magnetic field. A complete set
of transport coefficients, consistent with the Curie and Onsager
principles, is derived for thermal conduction, as well as
shear and bulk viscosities.
It is shown that in the most general case the dissipative function contains
five shear viscosities, two bulk viscosities, and three thermal conductivity
coefficients. We use Zubarev's non-equilibrium statistical operator method
to relate these transport coefficients to correlation functions of the
equilibrium theory.  The desired relations emerge at linear order
in the expansion of the non-equilibrium statistical operator
with respect to the gradients of relevant statistical
parameters (temperature, chemical potential, and velocity.) The transport
coefficients are cast in a form that can be conveniently computed using
equilibrium (imaginary-time) infrared Green's functions defined with
respect to the equilibrium statistical operator.
\end{abstract}

\end{frontmatter}

\section {Introduction}\label{introduction}

Computation of transport coefficients of relativistic systems is of
importance in a variety of fields, including relativistic astrophysics,
cosmology, and heavy-ion physics.  The kinetic theory based on the Boltzmann
equation provides an adequate tool for describing transport in these systems
in real time in the  dilute regime.  It may be systematically extended
to denser systems in cases where one can establish a hierarchy of the
timescales governing the system.  An alternative to the Boltzmann approach
is based on the Kubo method, in which non-equilibrium processes are regarded
as the response of the system to an external perturbation, which can be computed
from {\it equilibrium} correlation functions.  This method has the advantage
that the transport coefficients can be evaluated by using
equilibrium Green's functions formulated in imaginary time.

In this paper we shall adopt the correlation-function method developed by
Zubarev~\cite{Zubarev,Hosoya:1983id,Horsley:1985dz} to derive expressions
for transport coefficients of strongly magnetized relativistic fluids. The Zubarev
method is based on the concept of a non-equilibrium statistical operator, which
is a generalization of the equilibrium Gibbs statistical operator to
non-equilibrium states. This approach will enable us to obtain kinetic
transport coefficients in terms of correlation functions by expanding the
operator in terms of small gradients of the thermodynamical parameters. It
is generally assumed that the state of the system can be defined in terms of
(space-time dependent) fields of temperature, chemical potential, and momentum ---
a typical parameter set that is sufficient for a complete description of the system.
(The set of {\it relevant parameters} depends in general on the specific problem
under consideration.) The transport coefficients are then obtained from linear
perturbations of the non-equilibrium statistical operator around its equilibrium
value.  In this paper we will be concerned with the situation in which the
system is in an external Abelian gauge field, \ie, an electromagnetic field.

The physical motivation for studying relativistic fluids in strong magnetic
fields is provided by compact stellar objects (neutron stars).
It is now well established that neutron stars feature magnetic fields
of order $10^{12}$-$10^{13}$ G on average, with a subclass of neutron stars
(magnetars) having magnetic fields of order $10^{15}$-$10^{16}$ G.  Although
the {\it origin} of such large magnetic fields is not known, one possibility
is the conservation of magnetic flux~\cite{Woltjer,Ginzburg} under collapse
from ordinary stellar dimensions.  Alternatively, the field could be generated
by a dynamo mechanism~\cite{Thompson:1993hn}.  The field strengths $B\le 10^{16}$ G
inferred for magnetars~\cite{Kouveliotou:1998ze,Kouveliotou:1998fd} refer to the
surface of a compact star. The internal fields could be by several orders of
magnitude larger. An upper limit on the magnitude of the magnetic field
$B\le 10^{18}$~G is set by the virial theorem for a star in gravitational
equilibrium~\cite{Chandra,Astrophys:j383:745:1991}.  Observationally, soft gamma-ray
repeaters (SGRs) and anomalous x-ray pulsars (AXPs) are now identified with
magnetars; these objects have been associated with specific supernova remnants
and form a subclass of young strongly magnetized neutron stars which are distinct
from ordinary radio pulsars.

The {\it equilibrium} properties of dense matter in strong magnetic
fields have been studied intensively since the discovery of magnetars.
These properties include the equation of state of dense nuclear and quark
matter~\cite{astro-ph/0011148,astro-ph/9703034,astro-ph/9703066},
neutrino propagation and pulsar kicks, and
neutrino emission and cooling~\cite{Kusenko:1996sr,Horowitz:1997mk,
Arras:1998mv,Henley:2007mj,Maruyama:2010hu,Dicus:2007gb,Charbonneau:2009ax,vanDalen:2000zw}.
However, hydrodynamics and non-equilibrium processes in strong
fields have not been studied until recently.

In an earlier paper~\cite{Huang:2009ue}, we have developed an
anisotropic hydrodynamic theory to describe strongly magnetized
strange matter in a compact star. It was shown that the dissipative
processes in matter are completely described in terms of seven
viscosity coefficients (which include five shear viscosities and two
bulk viscosities), three thermal conductivities, and three
coefficients of electrical conductivity.  Our model is the
relativistic counterpart of the non-relativistic formulation of plasma
theory in strong magnetic fields due to Braginskii~\cite{Braginskii}.

The purpose of this work is to extend the description of the hydrodynamics
of strongly magnetized fluids by relating the hydrodynamic equations
to certain correlation functions. The latter can be computed using the
methods of equilibrium statistical mechanics. Among several methods to
derive Kubo formulas, we choose the one based on the idea of non-equilibrium
statistical operators (see Refs.~\cite{Zubarev,Hosoya:1983id,Horsley:1985dz}
for a detailed discussion of this method for non-magnetized fluids.)
Thus, compared to other relativistic treatments of
magnetohydrodynamics~(see, {\it e.~g.},
Refs.\ \cite{Gedalin:1991}), our approach relates the
transport coefficients entering magnetohydrodynamics to retarded
correlation functions of the microscopic theory.

This paper is organized as follows. Section~\ref{anisotropic}
is devoted to magnetohydrodynamics of relativistic fluids in
the ideal (Subsec.~\ref{ideal}) and dissipative (Subsec.~\ref{navier})
cases.  We derive a thermodynamically consistent set of equations
describing fluid dynamics in strong magnetic fields and identify
the full set of transport coefficients needed for a complete description
of dissipation.  In Sec.~\ref{nonequilibrium} we utilize the non-equilibrium
statistical-operator method in the formulation by Zubarev to derive Kubo
formulas relating the transport coefficients appearing in our
magnetohydrodynamic theory to correlation functions defined in terms
of the underlying elementary fields. A discussion of our results and
a comparison to previous work is given in Sec.~\ref{summary}.
Natural units $\hbar=k_B=c=1$ are adopted, and the SI system
of units is employed in equations involving electromagnetism.
The metric tensor is given by $g^{\m\n}=\diag(1,-1,-1,-1)$.
A number of calculational details are relegated to the appendices.

\section {Magnetohydrodynamics of relativistic fluids}\label{anisotropic}
\subsection {Non-dissipative fluids}\label{ideal}

Hydrodynamics arises as a long-wave-length, low-frequency
effective theory of matter from an expansion of the
energy-momentum tensor $T^{\m\n}$, the current(s) $n^\m$ of conserved
charge(s), and the entropy density flux $s^\m$, {\it etc.},
with respect to (small) gradients of the four-velocity $u^\m$ and
the thermodynamic parameters of the system, such as the
temperature $T$, the chemical potential $\m$, {\it etc.}
The zeroth-order terms in this expansion correspond to an ideal fluid,
which we will discuss here; the extension to the dissipative case
is given in the following subsection.

For the sake of simplicity, we will consider only
one conserved charge, {\it e.~g.}, the electric charge. Many physical
systems of interest may have other conserved charges, such as
baryon number, isospin, {\it etc}. The extension of our discussion
to multiple conserved charges is straightforward. The hydrodynamic
equations can be expressed as conservation laws for
the energy-momentum tensor $T^{\m\n}$ and the conserved currents, here
the electric current $n^\m$,
\begin{eqnarray}
\label{hydroeqn01}
\pt_\m T^{\m\n}&=&0,\\
\label{hydroeqn02}
 \pt_\m n^\m&=&0.
\end{eqnarray}
In the presence of an electromagnetic field the zeroth-order
contribution to the energy-momentum tensor can be written
as~\cite{Huang:2009ue,Groot:1969,Israel:1978up}
\begin{eqnarray}
\label{tmn00} T^{\m\n}_0&=&T_{\rm F0}^{\m\n}+T_{\rm EM}^{\m\n},\\
\label{tmn00_1}
T^{\m\n}_{\rm F0}&=&\ve\, u^\m u^\n-P\D^{\m\n}-\frac{1}{2}\lb
M^{\m\l}F^{\;\;\n}_{\l}+M^{\n\l}F^{\;\;\m}_{\l}\rb ,\\
\label{tmn00_2}
T_{\rm EM}^{\m\n}&=&-F^{\m\l}F_{\;\;\l}^\n+\frac{g^{\m\n}}{4}F^{\r\s}F_{\r\s},
\end{eqnarray}
where $F^{\m\n}$ is the field-strength tensor, $\ve$ and $P$ are
the local energy density and the thermodynamic pressure,
$\D^{\m\n}\equiv g^{\m\n}-u^\m u^\n$ is the projector on the
directions orthogonal to $u^\m$, and $M^{\m\n}$ is the polarization tensor.
In the non-dissipative limit, the entropy is conserved in addition to the charge,
\ie, it obeys a continuity equation analogous to Eq.\ (\ref{hydroeqn02}). The charge
and entropy currents in the non-dissipative theory are
\begin{eqnarray}
n_{0}^\m&=&n\, u^\m ,\\
s_0^\m &=& s\, u^\m ,
\end{eqnarray}
where $n$ and $s$ are the electric charge and entropy
densities measured in the rest frame of the fluid. For the following
discussion it will be convenient to decompose the tensor $F^{\m\n}$
into components parallel and perpendicular to $u^\m$,
\begin{eqnarray}
\label{fmn} F^{\m\n}&=&F^{\m\l}u_\l u^\n-F^{\n\l}u_\l
u^\m+\D^\m_{\;\;\a}F^{\a\b}\D_\b^{\;\;\n}\non &\equiv&E^\m u^\n-E^\n
u^\m+\frac{1}{2}\e^{\m\n\a\b}\lb u_\a B_\b-u_\b B_\a\rb,
\end{eqnarray}
where we have defined the four-vectors
$E^\m\equiv F^{\m\n}u_\n$ and $B^\m\equiv
\e^{\m\n\a\b}F_{\n\a}u_\b/2$, with $\e^{\m\n\a\b}$ denoting the totally
anti-symmetric Levi-Civita tensor. In the rest frame of the fluid,
$u^\m=(1,{\bf 0})$,  we find that  $E^0=B^0=0$, $E^i=F^{i0}$ and
$B^i=-\e^{ijk}F_{jk}/2$, where the indices $i, j, k$ run over 1, 2, 3.
The latter three vectors are precisely the electric and
magnetic fields in this frame. Therefore, $E^\m$ and $B^\m$ can be
interpreted as the electric and magnetic fields measured
in the frame in which the fluid moves with a velocity $u^\m$.

The antisymmetric polarization tensor $M^{\m\n}$ describes
the response of matter to the applied field strength $F^{\m\n}$. In
the case of systems described by a grand canonical ensemble, it is
given by $M^{\m\n}\equiv -\pt\O/\pt F_{\m\n}$ in terms of
the thermodynamic potential $\O$.  For later use, we also define
the in-medium field strength tensor $H^{\m\n}\equiv
F^{\m\n}-M^{\m\n}$. In analogy to what is done for $F^{\m\n}$, we can
decompose $M^{\m\n}$ and $H^{\m\n}$ as
\begin{eqnarray}
M^{\m\n}&=& P^\n u^\m-P^\m u^\n+\frac{1}{2}\e^{\m\n\a\b}\lb
M_\b u_\a-M_\a u_\b\rb, \\ H^{\m\n}&=& D^\m u^\n-D^\n
u^\m + \frac{1}{2}\e^{\m\n\a\b}\lb H_\b u_\a-H_\a u_\b\rb,
\end{eqnarray}
with $P^\m\equiv -M^{\m\n}u_\n$, $M^\m\equiv
\e^{\m\n\a\b}M_{\n\a}u_\b/2$,
$H^\m\equiv \e^{\m\n\a\b} H_{\n\a}u_\b/2$, and $D^\m\equiv H^{\m\n}u_\n$.

In the rest frame of the fluid, the non-trivial components of these
tensors are $(F^{10}, F^{20}, F^{30})=\bf{E}$, $(F^{32}, F^{13},
F^{21})=\bf{B}$, $(M^{10}, M^{20}, M^{30})=-\bf{P}$, $(M^{32},
M^{13}, M^{21})=\bf{M}$, $(H^{10}, H^{20}, H^{30})=\bf{D}$, and
$(H^{32}, H^{13}, H^{21})=\bf{H}$. Here $\bf{P}$ and $\bf{M}$ are
the electric polarization vector and magnetization vector,
respectively. In linear approximation they are related to the
fields $\bf{E}$ and $\bf{B}$ by the usual expressions
${\bf P}=\c_e{\bf E}$ and ${\bf
M}=\c_m{\bf B}$, where $\c_e$ and $\c_m$ are the electric and
magnetic susceptibilities. Note that the four-vectors $E^\m, B^\m, \cdots$ are
all space-like, \ie,  $E^\m u_\m=0, B^\m u_\m=0, \cdots$, and are normalized
as $E^\m E_\m=-E^2, B^\m B_\m=-B^2, \cdots$, with $E\equiv |\bf E|$
and $B\equiv |\bf B|$.

For many systems encountered in astrophysics, one example
being the interior of a neutron star, the electric field is much
weaker than the magnetic field.  Therefore, in the following discussion
we will omit the contributions originating from the electric field, but
will occasionally comment on how our expression will be
modified by it.  Next let us introduce the four-vector $b^\m\equiv
B^\m/B$, which is normalized as $b^\m b_\m=-1$ (in agreement with the
normalization of the vector $B^\m$ above), along with the antisymmetric rank-2 tensor
$b^{\m\n}\equiv\e^{\m\n\a\b} b_\a u_\b$. Then, upon dropping the electric field
in the field strength tensors, we write these in the compact form
\begin{eqnarray}
F^{\m\n}&=&-B b^{\m\n},\\ M^{\m\n}&=&-M b^{\m\n},\\
H^{\m\n}&=&-H b^{\m\n},
\end{eqnarray}
where $M\equiv |\bf M|$ and $H\equiv |\bf H|$. Again neglecting
electric fields, the matter and field contributions to
the energy-momentum tensor (\ref{tmn00}) can now be written
in terms of $b^{\mu}$ and $b^{\mu\nu}$
as~(see, {\it e.g.}, Refs.~\cite{Huang:2009ue,Gedalin:1991})
\begin{eqnarray}
\label{tmn0} T^{\m\n}_{\rm F0}&=&\ve u^\m
u^\n-\pperp\Xi^{\m\n}+\ppara b^\m b^\n,\\
T^{\m\n}_{\rm EM}&=&\frac{1}{2}B^2\lb u^\m u^\n-\Xi^{\m\n}-b^\m
b^\n\rb,
\end{eqnarray}
where $\Xi^{\m\n}\equiv\D^{\m\n}+b^\m b^\n$
is the projection tensor on the direction orthogonal to both
$u^\m$ and $b^\m$. Here we have defined the
transverse and longitudinal pressures $\pperp=P-MB$ and $\ppara=P$
relative to the vector $b^\m$. In the
absence of a magnetic field, the fluid is isotropic and
$\pperp=\ppara=P$, where $P$ is the thermodynamic pressure defined
in Eq.~(\ref{tmn00_2}). In the local rest frame of the fluid, we have
$b^\m=(0,0,0,1)$, where without loss of generality
the direction of the magnetic field is chosen as the $z$-axis.
We see then that the electromagnetic tensor takes the usual form,
while $T^{\m\n}_{\rm F0}=\diag(\ve,\pperp, \pperp, \ppara)$,
as required.

Next we would like to check the consistency of the terms that appear
in $T_{\rm F0}^{\m\n}$ with the formulas of standard thermodynamics
involving electromagnetic fields. By using the thermodynamic
relation
\begin{eqnarray}
\label{thermo} \ve&=&Ts+\m n-P
\end{eqnarray}
and the conservation equations for $n_{0}^\m$ and $s_0^\m$ in ideal
hydrodynamics, one can show that the hydrodynamic equation
$u_\n\pt_\m T^{\m\n}_0=0$ together with the Maxwell equation
(\ref{maxwell}) implies
\begin{eqnarray}
\label{de} D\ve=TDs+\m D n-MDB,
\end{eqnarray}
which is consistent with the standard thermodynamic relation
\begin{eqnarray}
\label{de2} d\ve&=&Tds+\m dn-MdB.
\end{eqnarray}
From \eq{thermo} and \eq{de2}, one can obtain the Gibbs-Duhem
relation,
\begin{eqnarray}
\label{gd} dP=sdT+nd\m+MdB.
\end{eqnarray}
We should note that the potential energy $-MB$ has already been
included in our definition of $\ve$. Otherwise, new terms $-MB$,
$-D(MB)$, and $-d(MB)$ should be added to the left-hand sides of
\eq{thermo}, \eq{de}, and \eq{de2}, respectively, while the
Gibbs-Duhem relation keeps the form of \eq{gd}.

The complete set of non-dissipative hydrodynamic equation
includes the Maxwell equations, which we state for completeness:
\begin{eqnarray}
\label{maxwell0} \pt_\n\lb B^\m u^\n-B^\n u^\m\rb&=&0,\\
\label{maxwell2}\pt_\m H^{\m\n}&=&n^\n.
\end{eqnarray}
Contracting Eq.~(\ref{maxwell0}) with $b_\m$ gives
the so-called induction equation
\begin{eqnarray}
\label{maxwell}\h+D\ln B-u^\n b^\m\pt_\m b_\n=0,
\end{eqnarray}
where $\h\equiv\pt_\m u^\m$ is the velocity divergence and $D\equiv
u^\m\pt_\m$ is the substantial (co-moving) time-derivative.
(Note that $b_{\mu}Db^{\mu} = b^{\mu}Db_{\mu}=0$.)
In closing this section, we confirm that the form of the
energy -momentum tensor $T_{\rm F0}^{\m\n}$ is consistent
with standard thermodynamic formulas for matter in
electromagnetic fields~\cite{Huang:2009ue}.

\subsection {Dissipative fluids}\label{navier}

The next-to-leading (first-order) derivative expansion of
conserved quantities leads to the Navier-Stokes-Fourier-Ohm
theory which, as is well known, includes dissipative effects.
The continuity equation for the entropy now changes to
\begin{equation}
T\pt_\m s^\m \geq 0
\end{equation}
(second law of thermodynamics).
In the dissipative theory the quantities $T^{\m\n}$,
$n^\m$, and $s^\m$ can be generally expressed as
\begin{eqnarray}
\label{tmn}
T^{\m\n}&=&T_0^{\m\n}+h^\m u^\n+h^\n u^\m+\t^{\m\n},\non
n^\m&=&nu^\m+j^\m,\non s^\m&=&s u^\m+j_s^\m,
\end{eqnarray}
where $h^\m$ is the energy flux, $\t^{\m\n}$ is the viscous stress tensor,
and $j^\m$ and $j_s^\m$ are the charge and entropy diffusion
fluxes. They are all orthogonal to $u^\m$, reflecting the fact
that the dissipation in the fluid should be spatial. We shall assume
that $j_s^\m$ can be expressed as a linear combination of $h^\m$ and
$j^\m$. This allows us to incorporate the fact that the entropy flux
is determined by the energy and charge diffusion
fluxes. Thus, following Refs.~\cite{Israel:1979wp,Hiscock:1985zz}, we write
\begin{eqnarray}
\label{energyeqn}
j_s^\m=\g h^\m-\a j^\m,
\end{eqnarray}
the coefficients $\g$ and $\a$ being functions of thermodynamic
variables.

In order to discuss the dissipative parts, let us first define the four-velocity
$u^\m$, since it is not unique when one allows for energy exchange by
thermal conduction.  We employ the Landau-Lifshitz frame, in
which $u^\m$ is chosen to be parallel to the energy density flow, \ie,
$u_{\nu}T^{\mu\nu}=u_{\mu} \varepsilon$.  It follows from the  first equation
(\ref{tmn}) that $h^\m=0$.  We project \eq{hydroeqn01} onto $u^\n$ and exploit the
fact that electric field $E_\l$ is zero.  Utilizing
Eq.~(\ref{energyeqn}), straightforward manipulations lead to
\begin{eqnarray}
\label{energyeqn2}
D\ve+(\ve+P)\h-\t^{\m\n}\pt_\m u_\n+MDB=j^\l
E_{\l}=0.
\end{eqnarray}
Combining this equation with the thermodynamic
relation $\epsilon = Ts+\mu n-P$ and the continuity equation
(\ref{hydroeqn02}), we arrive at
\begin{eqnarray}
\label{divs} T\pt_\s s^\s&=&\t^{\s\n}w_{\s\n}
-j^\s T\nabla_\s\a+(\m-T\a)\pt_\s j^\s,
\end{eqnarray}
where $\nabla_\s\equiv \D_{\s\n}\pt^\n$ and $w^{\s\n}\equiv
\frac{1}{2}\lb\nabla^\s u^\n+\nabla^\n u^\s\rb$.
The first and second terms on the right-hand side
of \eq{divs} (the dissipative function)
correspond to viscous and thermal dissipation, respectively, whereas
the last term vanishes due to our choice of $\alpha$.
For a thermodynamically and hydrodynamically stable system
the dissipative function should be non-negative.
This implies that for small perturbations it must be a
quadratic form, so we can write
\begin{eqnarray}
\label{linearcon1}
T\a&=&\m,\\
\label{linearcon2}
\t^{\m\n}&=&\w^{\m\n\a\b}w_{\a\b},\\
\label{linearcon3}
j^\m&=&\k^{\m\n}T\nabla_\n\a,
\end{eqnarray}
where $\w^{\m\n\a\b}$ is the rank-four tensor of
viscosity coefficients and $\k^{\m\n}$ is the thermal conductivity
tensor with respect to the diffusion flux of electric charge density.
By definition, $\w^{\m\n\a\b}$ is symmetric in
the pairs of indices $\a,\b$ and $\m,\n$. It necessarily satisfies
the condition $\w^{\m\n\a\b}(B^\s)=\w^{\a\b\m\n}(-B^\s)$, which is
Onsager's symmetry principle for transport coefficients. Similarly,
the tensors $\k^{\m\n}$ and $\s^{\m\n}$ should satisfy the
conditions $\k^{\m\n}(B^\l)=\k^{\n\m}(-B^\l)$ and
$\s^{\m\n}(B^\l)=\s^{\n\m}(-B^\l)$. Furthermore, all the tensors of
transport coefficients $\w^{\m\n\a\b}$, $\k^{\m\n}$, and $\s^{\m\n}$
must be orthogonal to $u^\m$ by definition.

The constructive equations connecting the irreversible
fluxes $\t_{\m\n}\,,j_\m$ and the thermodynamic forces $w_{\m\n}\,,
\nabla_\m\a$ in dissipative magnetohydrodynamics
may have a more general tensor structure
\begin{eqnarray}
\t^{\m\n}&=&\w^{\m\n\a\b}w_{\a\b}+\l^{\m\n\r}T\nabla_\r\a\;,\non
j^\m&=&\g^{\m\r\s}w_{\r\s}+\k^{\m\n}T\nabla_\n\a\;,
\end{eqnarray}
where $\l^{\m\n\r}$ and $\g^{\m\r\s}$ are rank-three tensors
representing the coupling between charge diffusion and
energy-momentum transport.  However, the Curie principle,
which states that linear transport relations can be realized
between irreducible tensors of the same {\it rank and parity,} does
not allow for such higher-rank tensors $\l^{\m\n\r}$ and
$\g^{\m\r\s}$. There are, however, some exceptions as, {\it e.~g.},
systems with parity violation~\cite{Kharzeev:2011ds}.
In strong magnetic fields the ranks of the tensors are changed compared
to the non-magnetic case because of the breaking of spatial symmetries
by the magnetic field.  However, the parity of the tensors with respect to
reflections should be preserved.  We therefore conclude that
$\l^{\m\n\r}=0=\g^{\r\m\n}$.

The tensors $\w^{\m\n\a\b}$ and $\k^{\m\n}$ can be decomposed into
sums of irreducible tensor combinations constructed
from the vectors $u^\m$, $b^\m$, $g^{\m\n}$, and $b^{\m\n}$.
In the case of the tensor $\w^{\m\n\a\b}$, all the independent
irreducible tensor combinations that share its symmetries
and are orthogonal to $u^\m$
are given by~\cite{Huang:2009ue}
\begin{eqnarray}
\label{combination} &&({\rm{i}})\quad \D^{\m\n}\D^{\a\b},\non
&&({\rm{ii}})\quad \D^{\m\a}\D^{\n\b}+\D^{\m\b}\D^{\n\a},\non
&&({\rm{iii}})\quad\D^{\m\n}b^\a b^\b+\D^{\a\b}b^\m b^\n,\non
&&({\rm{iv}})\quad b^\m b^\n b^\a b^\b,\non &&({\rm{v}})\quad
\D^{\m\a}b^\n b^\b+\D^{\n\b}b^\m b^\a+\D^{\m\b}b^\n
b^\a+\D^{\n\a}b^\m b^\b,\non &&({\rm{vi}})\quad \D^{\m\a}
b^{\n\b}+\D^{\n\b} b^{\m\a}+\D^{\m\b} b^{\n\a} +\D^{\n\a}
b^{\m\b},\non &&({\rm{vii}})\quad b^{\m\a}b^\n b^\b +b^{\n\b}b^\m
b^\a+b^{\m\b}b^\n b^\a+b^{\n\a}b^\m b^\b.
\end{eqnarray}
In Ref.~\cite{Huang:2009ue} the combination
$b^{\m\a}b^{\n\b}+b^{\m\b}b^{\n\a}$ was added to the above list.
However, as shown in Appendix~\ref{eight}, this
tensor can be written as a linear combination of tensors
listed in Eq.\ (\ref{combination}); specifically
$b^{\m\a}b^{\n\b}+b^{\m\b}b^{\n\a}=2
({\rm i}) - ({\rm ii})+2 ({\rm iii})-({\rm v})$.

The independent irreducible tensor combinations that have
the symmetry of $\k^{\m\n}$ and are orthogonal to $u^\m$
read
\begin{eqnarray}
\label{combination-k} ({\rm{viii}})\;\; \D^{\m\n},\quad
({\rm{ix}})\;\; b^\m b^\n,\quad ({\rm{x}})\;\;b^{\m\n}.
\end{eqnarray}
In order to separate the contributions of shear and bulk viscosities we will
need to construct tensor structures that differ from
those used in Ref.~\cite{Huang:2009ue}.
An appropriate
set is given by
\begin{eqnarray}
\label{combination2} &&({\rm{i'}})\quad
\D^{\m\n}\D^{\a\b}=({\rm{i}}),\non &&({\rm{ii'}})\quad
\D^{\m\a}\D^{\n\b}+\D^{\m\b}\D^{\n\a}=({\rm{ii}}),\non
&&({\rm{iii'}})\quad\Xi^{\m\n}\Xi^{\a\b}=2({\rm{iii}})+2({\rm{i}})+2({\rm{iv}}),\non
&&({\rm{iv'}})\quad b^\m b^\n b^\a b^\b=({\rm{iv}}),\non
&&({\rm{v'}})\quad \Xi^{\m\a}b^\n b^\b+\Xi^{\n\b}b^\m
b^\a+\Xi^{\m\b}b^\n b^\a+\Xi^{\n\a}b^\m
b^\b=({\rm{v}})+4({\rm{iv}}),\non &&({\rm{vi'}})\quad \Xi^{\m\a}
b^{\n\b}+\Xi^{\n\b} b^{\m\a}+\Xi^{\m\b} b^{\n\a} +\Xi^{\n\a}
b^{\m\b}=({\rm{vi}})+({\rm{vii}}),\non &&({\rm{vii'}})\quad
b^{\m\a}b^\n b^\b +b^{\n\b}b^\m b^\a+b^{\m\b}b^\n b^\a+b^{\n\a}b^\m
b^\b=({\rm{vii}}).
\end{eqnarray}
From these tensors we construct the following combinations, which
are either traceless,
\begin{eqnarray}
\label{traceless} &&({\rm{i''}})\quad
\D^{\m\a}\D^{\n\b}+\D^{\m\b}\D^{\n\a}-\frac{2}{3}\D^{\m\n}\D^{\a\b},\non
&&({\rm{ii''}})\quad
\lb\D^{\m\n}-\frac{3}{2}\Xi^{\m\n}\rb\lb\D^{\a\b}-\frac{3}{2}\Xi^{\a\b}\rb,\non
&&({\rm{iii''}})\quad -\Xi^{\m\a}b^\n b^\b-\Xi^{\n\b}b^\m
b^\a-\Xi^{\m\b}b^\n b^\a-\Xi^{\n\a}b^\m b^\b,\non
&&({\rm{iv''}})\quad -\Xi^{\m\a} b^{\n\b}-\Xi^{\n\b}
b^{\m\a}-\Xi^{\m\b} b^{\n\a} -\Xi^{\n\a} b^{\m\b},\non
&&({\rm{v''}})\quad b^{\m\a}b^\n b^\b +b^{\n\b}b^\m
b^\a+b^{\m\b}b^\n b^\a+b^{\n\a}b^\m b^\b,
\end{eqnarray}
or have non-vanishing trace,
\begin{eqnarray}
\label{tracepart}
&&({\rm{vi''}})\quad\frac{3}{2}\Xi^{\m\n}\Xi^{\a\b},\non
&&({\rm{vii''}})\quad 3b^\m b^\n b^\a b^\b.
\end{eqnarray}
Now we can write down the most general expression for $\w^{\m\n\a\b}$ as
\begin{eqnarray}
\label{etaexpression}
\w^{\m\n\a\b}=\w_0({\rm{i''}})+\w_1({\rm{ii''}})+\w_2({\rm{iii''}})+\w_3({\rm{iv''}})+\w_4({\rm{v''}})
+\zperp({\rm{vi''}})+\zpara({\rm{vii''}}).
\end{eqnarray}
Similarly, we also have the general decompositions for $\k^{\m\n}$, namely
\begin{eqnarray}
\label{kexpression} \k^{\m\n}=\k_\perp\Xi^{\m\n}-\k_\parallel b^\m
b^\n-\k_\times b^{\m\n}.
\end{eqnarray}
Thus we conclude that a fluid in a magnetic field has in general seven
independent viscosity coefficients and three independent thermal
conduction coefficients, which are defined through \eq{etaexpression}
and \eq{kexpression}, respectively.  Substituting
Eqs. (\ref{etaexpression}) and (\ref{kexpression}) into Eqs.~
(\ref{linearcon2}) and (\ref{linearcon3}), respectively, we obtain the
expressions for the viscous stress tensor and thermal flux in terms of
viscosity coefficients and thermal conductivities,
\begin{eqnarray}
\label{pmn}\t^{\m\n}&=&2\w_0\lb
w^{\m\n}-\frac{1}{3}\,\D^{\m\n}\h\rb+\w_1\lb\D^{\m\n}
-\frac{3}{2}\Xi^{\m\n}\rb\lb\h-\frac{3}{2}\f\rb\non
&-&2\w_2\lb
b^\m\Xi^{\n\a}b^\b+b^\n\Xi^{\m\a}b^\b\rb w_{\a\b}-2\w_3\lb
\Xi^{\m\a}b^{\n\b}+\Xi^{\n\a}b^{\m\b}\rb w_{\a\b}\non
&+&2\w_4\lb
b^{\m\a}b^\n b^\b+b^{\n\a}b^\m b^\b\rb w_{\a\b}
+\frac{3}{2}\zperp\Xi^{\m\n}\f+3\zpara b^\m b^\n \vf\;,
\end{eqnarray}
\begin{eqnarray}
\label{jbm} j^\m&=&\k_\perp T\Xi^{\m\n}\nabla^\n\a-\k_\parallel b^\m
b^\n T\nabla_\n\a -\k_\times b^{\m\n} T\nabla_\n\a\;,
\end{eqnarray}
where $\f\equiv \Xi^{\m\n}w_{\m\n}$ and $\vf\equiv b^\m b^\n w_{\m\n}$.
The coefficients of the tensor $\t^{\m\n}$ multiplying
traceless tensors are identified as shear viscosities, while the
coefficients multiplying the tensors with non-vanishing trace are identified as
bulk viscosities.  The $\k$'s are the thermal conductivities.

The divergence of the entropy-density flux (\ref{divs}) can now be
written explicitly as
\begin{eqnarray}
\label{divs2} T\pt_\m s^\m&&\!\!\!\!\!\!=2\w_0\lb
w^{\m\n}-\frac{1}{3}\D^{\m\n}\h\rb\lb
w_{\m\n}-\frac{1}{3}\D_{\m\n}\h\rb+\w_1\lb\h-\frac{3}{2}\f\rb^2\non
&+&2\w_2\lb b^\m b_\r w^{\r\n}-b^\n b_\r w^{\r\m}\rb\lb b_\m b^\r
w_{\r\n}-b_\n b^\r w_{\r\m}\rb\non
&+&\frac{3}{2}\zperp\f^2+3\zpara\vf^2-\k_\perp
T^2\Xi^{\m\n}\nabla_\n\a\nabla_\m\a+\k_\parallel T^2\lb
b^\m\nabla_\m\a\rb^2.
\end{eqnarray}
One should note that the terms corresponding to the transport
coefficients $\w_3,\w_4$, and $\k_\times$ in \eqs{pmn}{jbm} do not
contribute to the divergence of the entropy-density flux. For stable
systems, all the other transport coefficients must be positive
to satisfy the second law of thermodynamics (positivity of the dissipative
function.) There are, however, examples in which this condition may
be violated. Indeed, as shown in Ref.~\cite{Huang:2009ue},
strange quark matter (if it occurs in some form in compact stars)
may have negative transverse bulk viscosity $\zperp$ in strong
magnetic fields, in situations where non-leptonic weak processes are
dominant.  The change in the sign of this transport coefficient can be
linked to the onset of a hydrodynamic instability~\cite{Huang:2009ue}.

Having determined the general form of the relativistic hydrodynamic
equations in strong magnetic fields, we turn now to the problem of
determining the transport coefficients entering these equations.
We have previously determined the bulk viscosity~\cite{Huang:2009ue}
in the specific case of strange matter, where knowledge of
the equation of state and perturbative weak interactions
rates is sufficient to fix the value of $\z$.
In order to obtain the remaining transport coefficients one must resort
to the microscopic theory. Two general approaches are available for this
task: one is based on the solution of a (linearized) Boltzmann equation;
the second is based on the evaluation of certain correlation functions.
The first approach is suitable for dilute systems or systems with
well-defined quasiparticles. The second approach is more general, since
the required correlation functions can in principle be evaluated for
a system with arbitrarily strong interactions  far from equilibrium.
We will follow the second approach to express the transport coefficients
of a fluid in a strong magnetic field in terms correlation functions.
In order to do so, we apply the powerful method of non-equilibrium statistical
operators developed by Zubarev~\cite{Zubarev}.

\section {Non-equilibrium Statistical Operators and Kubo Formulas}\label{nonequilibrium}

Zubarev's method of non-equilibrium statistical operators (NESO) allows one
to develop a statistical theory of irreversible processes via
a generalization of the Gibbs approach to equilibrium systems.  Zubarev's
method assumes that one can construct a statistical ensemble
representing the macroscopic state of the system in non-equilibrium.
In general this is possible if one is interested in quantities that are
averaged over time intervals that are long enough that the initial
state of the system is unimportant and the number of parameters
describing the system is correspondingly reduced.

We consider the system to be in the
hydrodynamic regime (in which the thermodynamical parameters such as
temperature and chemical potentials can be defined locally) and,
following Zubarev, choose the NESO in the form~\cite{Zubarev}
\begin{eqnarray}
\label{neso1} \rh(t)&=&Q^{-1}\exp{\ls-\int d^3\bx\, \zh(\bx,t)\rs},\\
Q&=&\Tr\exp{\ls-\int d^3\bx\, \zh(\bx,t)\rs},
\end{eqnarray}
where the operator $\zh$ reads
\begin{eqnarray}
\label{b1} \zh(\bx,t)&=&\ve\int_{-\infty}^t dt_1 e^{\ve(t_1-t)}\ls
\b^\n(\bx,t_1) \th_{0\n}(\bx,t_1)-\a(\bx,t_1) \nh^0(\bx,t_1)\rs,
\end{eqnarray}
with $\ve\ra+0$ after the thermodynamic limit is taken, and
\begin{eqnarray}
\b^\n(\bx,t)&=&\b(\bx,t) u^\n(\bx,t),\\ \a(\bx,t)&=&\b(\bx,t)
\m(\bx,t).
\end{eqnarray}
Physically the parameters $\b$, $\m$, and $u^\m$ stand for the
inverse local equilibrium temperature, chemical potential, and
flow four-velocity, respectively.  The conditions needed to make
such identifications will be addressed below.
The operators of
the energy-momentum tensor $\th^{\m\n}$ and the electric current
$\nh^\m$ satisfy the local conservation laws
\begin{eqnarray}
\label{conserve} \pt_\m\th^{\m\n}=0\;,\;\; \pt_\m \nh^\m=0.
\end{eqnarray}
Integrating \eq{b1} by parts
and using the conservation laws (\ref{conserve}),
we rewrite the exponent of Eq.~(\ref{neso1}) as
\begin{eqnarray}
\label{b3} \int d^3\bx\,\zh(\bx,t)&=& \int d^3\bx\ls \b^\n(\bx,t)
\th_{0\n}(\bx,t)-\a(\bx,t) \nh^0(\bx,t)\rs\non\!\!\!&-&\!\!\int
d^3\bx\int_{-\infty}^t dt_1 e^{\ve(t_1-t)}\Big[
\th_{\m\n}(\bx,t_1)\pt^\m
\b^\n(\bx,t_1)\non
&-&\nh^\m(\bx,t_1)\pt_\m\a(\bx,t_1)\Big].
\end{eqnarray}
The quantities $\pt^\m \b^\n, \pt_\m\a$ are thermodynamic ``forces''
as they involve gradients of velocity field, temperature, and
chemical potential. It is natural to identify the first integral
in Eq.~(\ref{b3}) as the equilibrium part of the statistical
operator and the second as the non-equilibrium part of the
statistical operator (the latter vanishing when the space-time
variations of the parameters $\a(\bx,t)$ and $\b^\n(\bx,t)$ vanish.)
Thus we may define the {\it local equilibrium part} of the
statistical operator as
\begin{eqnarray}
\label{eso1} \rh_l(t)&=&Q_l^{-1}\exp{\ls-\int d^3\bx \lb\b^\n(\bx,t)
\th_{0\n}(\bx,t)-\a(\bx,t) \nh^0(\bx,t)\rb\rs},\\
Q_l&=&\Tr\exp{\ls-\int d^3\bx \lb\b^\n(\bx,t)
\th_{0\n}(\bx,t)-\a(\bx,t) \nh^0(\bx,t)\rb\rs},
\end{eqnarray}
and write the complete statistical operator as
\begin{eqnarray}
\label{neso2} \rh &=&Q^{-1}e^{-\ah+\bh},\\ Q&=&\Tr e^{-\ah+\bh},
\end{eqnarray}
having introduced the following short-hand notation for the integrals
\begin{eqnarray}
\label{conv_a}
\ah(t)&=&\int d^3\bx \lb\b^\n(\bx,t)
\th_{0\n}(\bx,t)-\a(\bx,t) \nh^0(\bx,t)\rb,\\
\label{conv_b}
\bh(t)&=&\int
d^3\bx\int_{-\infty}^t dt_1 e^{\ve(t_1-t)}\ch(\bx,t_1),
\end{eqnarray}
with
\begin{eqnarray}
\ch(\bx,t)=\th_{\m\n}(\bx,t)\pt^\m
\b^\n(\bx,t)-\nh^\m(\bx,t)\pt_\m\a(\bx,t).
\end{eqnarray}

Now we consider the case of small perturbations from the equilibrium
state, \ie, the case in which the thermodynamic forces are sufficiently
small that the time-scales for relaxation of non-equilibrium states are
large.  In this regime, the relations between the thermodynamic forces and
irreversible currents are linear, so that we can approximate the
non-equilibrium statistical operator by keeping only the linear term
in the expansion of the exponent in Eq.~(\ref{neso2}). Thus, in the
linear approximation the complete statistical operator reads
\begin{eqnarray}
\label{neso3} \rh&=&\ls1+\int_0^1 d\t \lb e^{-\ah\t}\bh
e^{\ah\t}-\lan\bh\ran_l\rb\rs\rh_l,
\end{eqnarray}
where $\lan\bh\ran_l=\Tr\lb\rh_l\bh\rb$ is the average over the
local equilibrium operator.

Now we are in a position to evaluate the energy-momentum tensor
averaged over the non-equilibrium distribution. We obtain
\begin{eqnarray}
\lan
\th^{\m\n}(\bx,t)\ran&\equiv&\Tr\ls\r(t)\th^{\m\n}(\bx,t)\rs\non&=&\lan
\th^{\m\n}(\bx,t)\ran_l+\d\lan\th^{\m\n}(\bx,t)\ran,
\end{eqnarray}
where
\begin{eqnarray}
\label{T_perturb}
\d\lan \th^{\m\n}(\bx,t)\ran&=&\Tr\ls\int_0^1 d\t e^{-\ah\t}\bh
e^{\ah\t}\rh_l(t)\th^{\m\n}(\bx,t)\rs\non
&-&\Tr\ls\lan\bh\ran_l\rh_l(t)
\th^{\m\n}(\bx,t)\rs
\end{eqnarray}
is the deviation from the average equilibrium value
of the energy-momentum tensor $\lan \th^{\m\n}(\bx,t)\ran_l$.
Substituting the integral (\ref{conv_b}) into Eq.~(\ref{T_perturb}),
we obtain
\begin{eqnarray}
\d\lan \th^{\m\n}(\bx,t)\ran
&=&\int d^3\bx_1\int_{-\infty}^t dt_1 e^{\ve(t_1-t)}\non
&&\int_0^1\!\!
d\t\lan\th^{\m\n}(\bx,t)\ls e^{-\ah\t}\ch(\bx_1,t_1)
e^{\ah\t}-\lan\ch(\bx_1,t_1)\ran_l\rs\ran_l\;.
\end{eqnarray}
The notation can be compactified by introducing
Kubo correlation functions defined as
\begin{eqnarray}
\lb{\hat X}(\bx,t),{\hat Y}(\bx',t')\rb\equiv\int_0^1 d\t\lan {\hat
X}(\bx,t)\ls e^{-\ah\t}{\hat Y}(\bx',t')e^{\ah\t}-\lan{\hat
Y}(\bx',t')\ran_l\rs\ran_l,
\end{eqnarray}
in terms of which we finally  obtain
\begin{eqnarray}
\label{averaget}
\tau^{\mu\nu} = \d\lan \th^{\m\n}(\bx,t)\ran&=&\int
d^3\bx_1\int_{-\infty}^t dt_1 e^{\ve(t_1-t)}\lb \th^{\m\n}(\bx,t),
\ch(\bx_1,t_1)\rb.
\end{eqnarray}
Similarly, the linear response of the charge current may be written as
\begin{eqnarray}
\label{averagen}
j^{\mu}= \d\lan \nh^{\m}(\bx,t)\ran&=&\int
d^3\bx_1\int_{-\infty}^t dt_1 e^{\ve(t_1-t)}\lb \nh^{\m}(\bx,t),
\ch(\bx_1,t_1)\rb.
\end{eqnarray}
Equations (\ref{averaget}) and (\ref{averagen}) describe the response
of the fluid to small perturbations.

In the next step we need to match the averages of the conserved quantities
(\ref{averaget}) and (\ref{averagen}) to those that appear in the
hydrodynamic equations. For this purpose, we must first specify the parameters
$\b$ and $\m$ as the inverse temperature and the
chemical potential in local equilibrium, respectively. In order
to enforce such matching it is
sufficient to require that in the rest frame
of the fluid~\cite{Zubarev,Horsley:1985dz,Mori:1958zz,Chapman},
\begin{eqnarray}
\label{matchb1}
\lan\th^{00}(\bx,t)\ran&=&\lan\th^{00}(\bx,t)\ran_l,\\
\label{matchb2}
\lan\nh^{0}(\bx,t)\ran&=&\lan\nh^{0}(\bx,t)\ran_l,
\end{eqnarray}
or in a covariant form $u^\m\d\lan\th_{\m\n}\ran u^\n=0$ and
$u^\m\d\lan\nh_{\m}\ran=0$.
Secondly, we need to identify the parameter $u^\m$ with the
four-velocity appearing in our hydrodynamic equations in
Sec.~\ref{anisotropic}.
In general, such identification requires a specific frame,
\eg, the Landau-Lifshitz (LL) frame, where the four-velocity
is parallel to the energy flow, or the Eckart frame where the
four-velocity is parallel to the charge flow.  From
\eqs{averaget}{averagen} we see that in the frame where $v^i=0$,
neither $\d\lan\th^{0i}\ran$ nor $\d\lan\nh^{i}\ran$ necessarily
vanishes, \ie, we require a transformation to either the
LL or Eckart frame.  We choose to work in the LL frame,
in which case the three-velocity should be boosted as
$v^i=\d\lan\th^{0i}\ran/(\ve+P)$, where
$\ve=\lan\th^{00}\ran_l,\,P=\sum_i\lan\th^{ii}\ran_l/3$, and
$n=\lan\nh^0\ran_l$ are the equilibrium values of energy density,
pressure, and charge density, respectively. (In the Eckart frame
the boost is given by $v^i=\d\lan\nh^{i}\ran/n$.)

Written in the LL frame, the hydrodynamic stress tensor and electric
current are~\cite{Horsley:1985dz}
\begin{eqnarray}
\label{match} T^{\m\n}&=&\lan\th^{\m\n}\ran_l+\d\lan
\th^{\m\n}\ran-u^\m u^\r\d\lan\th_{\r\s}\ran\D^{\s\n}-u^\n
u^\r\d\lan\th_{\r\s}\ran\D^{\s\m},\\
n^\m&=&\lan\nh^\m\ran_l+\d\lan
\nh^{\m}\ran-\frac{n}{\ve+P}u^\r\d\lan\th_{\r\s}\ran\D^{\s\m},
\end{eqnarray}
where the quantities on the left are those appearing in the hydrodynamic
discussion of Sec.~\ref{anisotropic}.  According to Eqs.\
(\ref{matchb1}) and (\ref{matchb2})
we have $u^\m\d\lan\th_{\m\n}\ran u^\n =0$ and $u^\m\d\lan\nh_{\m}\ran = 0$,
so any linear combination of these expressions can be freely added to the
right-hand side of \eq{match}.  This allows us to rewrite $T^{\m\n}$ and
$n^\m$ as
\begin{eqnarray}
\label{match2_1}
T^{\m\n}&=&\lan\th^{\m\n}\ran_l+\d\lan\kh^{\m\n}\ran,\\
\label{match2_2}
n^\m&=&\lan\nh^\m\ran_l-\d\lan \gh^{\m}\ran
\end{eqnarray}
after introducing
\begin{eqnarray}
\kh_{\m\n}&=&\th^{\r\s}\lb\D_{\r\m}\D_{\s\n}+\H_\b u_\r
u_\s\D_{\m\n}+\F_\b u_\r u_\s\Xi_{\m\n}\rb\non
&+&\nh^\r
u_\r\lb\H_\a\D_{\m\n}+\F_\a\Xi_{\m\n}\rb,\\
\gh_\m&=&\frac{n}{\ve+P}u_\r
\th^{\r\n}\D_{\n\m}-\nh^\n\D_{\n\m},
\end{eqnarray}
where $\H_\b$, $\H_\a$, $\F_\b$, and $\F_\a$ are defined in~\ref{derivation}.

The advantage of casting $\th^{\m\n}$ and $\nh^\m$ in the form given by
Eqs.~(\ref{match2_1}) and (\ref{match2_2}) is that the operator $\hat C$
can be decomposed as (see ~\ref{derivation})
\begin{eqnarray}
\label{c} \ch&=&\gh_\r\nabla^\r\a+\b\kh_{\r\s} w^{\r\s},
\end{eqnarray}
where $\gh_\r$ has odd spatial parity and
$\kh_{\r\s}$ has even spatial parity. Later it will be seen that such
a decomposition permits us to write the transport coefficients in a
symmetric manner.

For further convenience, we decompose
the energy-momentum tensor $\th^{\m\n}$ and the current $\nh^\m$
as follows:
\begin{eqnarray}
\th^{\m\n}&=&\hat{\ve}u^\m
u^\n-\hat{P}_\perp\Xi^{\m\n}+\hat{P}_\parallel b^\m b^\n+\hat{h}^\m
u^\n+\hat{h}^\n u^\m-\hat{R}^\m b^\n-\hat{R}^\n
b^\m+\hat{\p}^{\m\n},\non \nh^\m&=&\hat{n} u^\m+\hat{J}^\m+\hat{l}
b^\m,
\end{eqnarray}
the expansion coefficients being defined by
\begin{eqnarray}
\hat{\ve}&\equiv& u_\m u_\n \th^{\m\n},\non
\hat{P}_\perp&\equiv&-\frac{1}{2}\Xi_{\m\n}\th^{\m\n},\non
\hat{P}_\parallel&\equiv& b_\m b_\n\th^{\m\n},\non \hat{h}^\m
&\equiv&\D^{\m\r}\th_{\r\s}u^\s,\non
\hat{R}^\m&\equiv&\Xi^{\m\r}\th_{\r\s}b^\s,\non
\hat{\p}^{\m\n}&\equiv&\lb\Xi^{\r\m}\Xi^{\s\n}-\frac{1}{2}\Xi^{\r\s}\Xi^{\m\n}\rb\th_{\r\s},\non
\hat{n}&\equiv& \nh^\m u_\m,\non
\hat{J}^\m&\equiv&\nh_\n\Xi^{\m\n},\non \hat{l}&\equiv& -\nh^\m
b_\m.
\end{eqnarray}
The operators $\hat{h}^\m, \hat{R}^\m, \hat{\p}^{\m\n}$, and
$\hat{J}^\m$ are all perpendicular to $u_\m$, while $\hat{R}^\m,
\hat{\p}^{\m\n}$, and $\hat{J}^\m$ are also perpendicular to $b_\m$.
The physical meanings of these operators are obvious. With these
definitions we finally obtain
\begin{eqnarray}
\label{eq:k1}
\kh_{\m\n}&=&\hat{\p}^{\m\n}-\tilde{P}_\perp\Xi^{\m\n}+\tilde{P}_\parallel
b^\m b^\n-\hat{R}^\m b^\n-\hat{R}^\n b^\m,\\
\label{eq:g1}
\gh_\m&=&\frac{n}{\ve+P}h_\m-\hat{J}_\m-\hat{l} b_\m,
\end{eqnarray}
with
$\tilde{P}_\perp\equiv\hat{P}_\perp-(\H_\b+\F_\b)\hat{\ve}-(\H_\a+\F_\a)\hat{n}$,
and
$\tilde{P}_\parallel\equiv\hat{P}_\parallel-\H_\b\hat{\ve}-\H_\a\hat{n}$.

The dissipative terms can now be written as correlation functions constructed
from Eqs.~(\ref{eq:k1}) and (\ref{eq:g1}),
\begin{eqnarray}
\label{dtmn}
\t^{\m\n}(\bx,t)&=&\d\lan\th^{\m\n}(\bx,t)\ran=\b\int
d^3\bx_1\int_{-\infty}^t dt_1 e^{\ve(t_1-t)}\non
&\times&\lb \kh_{\m\n}(\bx,t),
\kh^{\r\s}(\bx_1,t_1)\rb w_{\r\s}(\bx_1,t_1),\\ \label{dnm}
j^{\m}(\bx,t)&=&\d\lan\nh^{\m}(\bx,t)\ran\non&=&-\int
d^3\bx_1\int_{-\infty}^t dt_1 e^{\ve(t_1-t)}\lb \gh^{\m}(\bx,t),
\gh^\r(\bx_1,t_1)\rb\nabla_\r\a(\bx_1,t_1),\non
\end{eqnarray}
where we have invoked Curie's principle that the correlators
between operators with different spatial parity must vanish.

Now suppose that the changes of the thermodynamical forces are
sufficiently small over the correlation lengths of these expressions,
such that they can be factored out from the integral. Then we obtain the
desired linear relations between the thermodynamical forces and the
irreversible currents. The coefficients in these linear relations
are easily identified as the viscosities and thermal conductivities:
\begin{eqnarray}
\label{kubovis} \w^{\m\n\r\s}&=&\b\int d^3\bx_1\int_{-\infty}^t dt_1
e^{\ve(t_1-t)}\lb \kh^{\m\n}(\bx,t), \kh^{\r\s}(\bx_1,t_1)\rb,\\
\label{kuboheat} \k^{\m\n}&=&-\b\int d^3\bx_1\int_{-\infty}^t dt_1
e^{\ve(t_1-t)}\lb \gh^{\m}(\bx,t), \gh^\n(\bx_1,t_1)\rb.
\end{eqnarray}
For practical purposes, it is more convenient to express these
transport coefficients in terms of retarded Green's functions,
which can be computed using the methods of equilibrium Green's functions.
Straightforward manipulations lead us (see Appendix~\ref{kubogreen} for
details) to
\begin{eqnarray}
\label{kubovis2}
\w^{\m\n\r\s}&=&i\frac{\pt}{\pt\o}
G_\w^{\m\n\r\s}({\bf 0},\o)\Big|_{\o\ra0},\\
\label{kuboheat2}
\k^{\m\n}&=&-i\frac{\pt}{\pt\o}
G_\k^{\m\n}({\bf0},\o)\Big|_{\o\ra0},
\end{eqnarray}
where $G_\w^{\m\n\r\s}(\bk,\o)$ and $G_\k^{\m\n}(\bk,\o)$ are the
Fourier transforms of the retarded Green's functions
\begin{eqnarray}
G_\w^{\m\n\r\s}(\bx,t)&\equiv&-i\h(t)\ls\kh^{\m\n}(\bx,t),\kh^{\r\s}({\bf
0},0)\rs,\non
G_\k^{\m\n}(\bx,t)&\equiv&-i\h(t)\ls\gh^{\m}(\bx,t),\gh^{\n}({\bf
0},0)\rs.
\end{eqnarray}
(Note that we are using the same notation for Green's functions in
real space-time and for their Fourier transforms in momentum-energy space.)
Next, we match \eq{kubovis2} with \eq{etaexpression}, which leads us to
\begin{eqnarray}
\D_{\m\n}\w^{\m\n\a\b}\Xi_{\a\b}&=&6\zperp,\non
\D_{\m\n}\w^{\m\n\a\b}b_\a b_\b&=&-3\zpara,\non
\D_{\m\a}\w^{\m\n\a\b}\D_{\n\b}&=&10\w_0+\frac{3}{2}\w_1+4\w_2+3\zperp+3\zpara,\non
\D_{\m\a}\w^{\m\n\a\b}b_\n
b_\b&=&-\frac{10}{3}\w_0-\w_1-2\w_2-3\zpara,\non b_\m b_\n b_\a
b_\b\w^{\m\n\a\b}&=&\frac{4}{3}\w_0+\w_1+3\zpara,\non
b_{\m\a}\Xi_{\n\b}\w^{\m\n\a\b}&=&-8\w_3,\non b_{\m\a} b_\n
b_\b\w^{\m\n\a\b}&=&2\w_4.
\end{eqnarray}
These equations uniquely determine the viscosity coefficients; explicitly,
they are given by
\begin{eqnarray}
\label{kubovis3}
\zperp&=&\frac{1}{6}\D_{\m\n}\w^{\m\n\a\b}\Xi_{\a\b},\non
\zpara&=&-\frac{1}{3}\D_{\m\n}\w^{\m\n\a\b}b_\a b_\b,\non
\w_0&=&\frac{1}{8}\lb-\D_{\m\n}\D_{\a\b}-2\D_{\m\n} b_\a
b_\b+2\D_{\m\a}\D_{\n\b}+4\D_{\m\a}b_\n b_\b+b_\m b_\n b_\a
b_\b\rb\w^{\m\n\a\b},\non
\w_1&=&\frac{1}{6}\lb\D_{\m\n}\D_{\a\b}+8\D_{\m\n}b_\a
b_\b-2\D_{\m\a}\D_{\n\b} - 4 \D_{\m\a}b_\n b_\b + 5 b_\m b_\n b_\a
b_\b\rb\w^{\m\n\a\b},\non
\w_2&=&\frac{1}{8}\lb\D_{\m\n}\D_{\a\b}+2\D_{\m\n} b_\a
b_\b-2\D_{\m\a}\D_{\n\b} - 8 \D_{\m\a}b_\n b_\b - 5 b_\m b_\n b_\a
b_\b\rb\w^{\m\n\a\b},\non
\w_3&=&-\frac{1}{8}b_{\m\a}\Xi_{\n\b}\w^{\m\n\a\b},\non
\w_4&=&\frac{1}{2}b_{\m\a} b_\n b_\b\w^{\m\n\a\b}.
\end{eqnarray}
We obtain the thermal conductivities in similar fashion by matching
\eq{kuboheat2} with \eq{kexpression}. Their explicit form is given by
\begin{eqnarray}
\label{kuboheat3}
\k_\perp&=&\frac{1}{2}\Xi_{\m\s}\k^{\m\s},\non
\k_\parallel&=&-b_\m
b_\s\k^{\m\s},\non\k_\times&=&-\frac{1}{2}b_{\m\s}\k^{\m\s}.
\end{eqnarray}
Finally, we wish to express the transport coefficients in terms of
the Green's functions. Substituting the results for
the viscosity and thermal conductivity tensors,
Eqs.~(\ref{kubovis2}) and (\ref{kuboheat2}), into Eqs.~(\ref{kubovis3})
and (\ref{kuboheat3}), we arrive at
\begin{eqnarray}
\label{kubovis4} \zperp&=&-\frac{1}{3}\frac{\pt}{\pt\o}\ls 2
G_{\tilde{P}_\perp\tilde{P}_\perp}^R({\bf
0},\o)+G_{\tilde{P}_\parallel\tilde{P}_\perp}^R({\bf
0},\o)\rs\Big|_{\o\ra0},\non \zpara&=&-\frac{1}{3}\frac{\pt}{\pt\o}
\ls 2 G_{\tilde{P}_\perp\tilde{P}_\parallel}^R({\bf
0},\o)+G_{\tilde{P}_\parallel\tilde{P}_\parallel}^R({\bf
0},\o)\rs\Big|_{\o\ra0},\non \w_0&=&-\frac{1}{4}\frac{\pt}{\pt\o}\im
G_{\hat{\p}_{\m\n},\hat{\p}^{\m\n}}^R({\bf 0},\o)\Big|_{\o\ra0},\non
\w_1&=&-\frac{4}{3}\w_0+2\frac{\pt}{\pt\o}
G_{\tilde{P}_\parallel\tilde{P}_\perp}^R({\bf
0},\o)\Big|_{\o\ra0},\non
\w_2&=&-\w_0+\frac{1}{2}\frac{\pt}{\pt\o}\im
G_{\hat{R}_\m,\hat{R}^\m}^R({\bf 0},\o)\Big|_{\o\ra0},\non
\w_3&=&\frac{1}{8}\frac{\pt}{\pt\o}
G_{b_{\r\a}\hat{\p}^{\r\s},\hat{\p}^\a_{\s}}^R({\bf
0},\o)\Big|_{\o\ra0},\non\w_4&=&-\frac{1}{2}\frac{\pt}{\pt\o}
G_{b_{\r\s}\hat{R}^{\r},\hat{R}^\s}^R({\bf 0},\o)\Big|_{\o\ra0},
\end{eqnarray}
where the retarded Green's function is defined as
$$G_{\ah\bh}^R(\bx,t)\equiv-i\h(t)\ls\ah(\bx,t),\bh({\bf0},0)\rs .$$
In deriving Eq.~(\ref{kubovis4}), we have used the properties
$-\im G_{\ah\ah}^R(\o,b^\s)=\im G_{\ah\ah}^R(-\o,b^\s)$ and
$G_{\ah\bh}^R(\o,b^\s)=\e_A\e_B G_{\bh\ah}^R(\o,-b^\s)$
of the retarded Green's function, which are valid for any Hermitian
bosonic operators $\ah$ and $\bh$.  (Here $\e_A$ ($\e_B$)
is the parity of the operator $\ah$ ($\bh$) under time-reversal.) These
relations are the consequence of the Onsager's reciprocal principle~\cite{Zubarev}.
Similarly, for the thermal conductivities we have
\begin{eqnarray}
\label{kuboheat4} \k_\perp&=&\frac{1}{2}\frac{\pt}{\pt\o}\im
G^R_{\Xi_{\m\n} \gh^\m,\gh^\n}({\bf 0},\o)\Big|_{\o\ra0},\non
\k_\parallel&=&-\frac{\pt}{\pt\o}\im G^R_{b_\m
\gh^\m,b_\n\gh^\n}({\bf 0},\o)\Big|_{\o\ra0},\non
\k_\times&=&-\frac{1}{2}\frac{\pt}{\pt\o} G^R_{b_{\m\n}
\gh^\m,\gh^\n}({\bf 0},\o)\Big|_{\o\ra0}.
\end{eqnarray}
For practical purposes it is sufficient to compute the transport coefficients
in the rest frame of the fluid.  In this case the correlation functions
for the $\eta$'s and $\kappa$'s simplify, and we find
\begin{eqnarray}
\label{restkubo} \w_0&=&-\frac{\pt}{\pt\o}\im
G^R_{\th^{12},\th^{12}}({\bf 0},\o)\Big|_{\o\ra0},\non
\w_2&=&-\w_0-\frac{\pt}{\pt\o}\im G^R_{\th^{13},\th^{13}}({\bf
0},\o)\Big|_{\o\ra0},\non \w_3&=&-\frac{1}{2}\frac{\pt}{\pt\o}
G_{\tilde{P}_\perp,\hat{T}^{12}}^R({\bf
0},\o)\Big|_{\o\ra0},\non\w_4&=&-\frac{\pt}{\pt\o}
G_{\hat{T}^{13},\hat{T}^{23}}^R({\bf 0},\o)\Big|_{\o\ra0},\non
\k_\perp&=&-\frac{\pt}{\pt\o}\im G^R_{\gh^1,\gh^1}({\bf
0},\o)\Big|_{\o\ra0},\non \k_\parallel&=&-\frac{\pt}{\pt\o}\im
G^R_{\gh^3,\gh^3}({\bf 0},\o)\Big|_{\o\ra0},\non
\k_\times&=&-\frac{\pt}{\pt\o} G^R_{\gh^1,\gh^2}({\bf
0},\o)\Big|_{\o\ra0}.
\end{eqnarray}
Equations (\ref{kubovis3})-(\ref{restkubo}) are our main results.
They express the transport coefficients of strongly magnetized fluids in terms
of equilibrium (retarded) correlation functions, which can be computed
by methods of equilibrium statistical mechanics. They take into account the
fact that the magnetic field breaks the translational symmetry of the problem,
thereby increasing the number of transport coefficients compared to the
nonmagnetic case. These expressions are valid at large field strengths (but below
values at which vacuum quantum fluctuations become important) and can be used
to compute transport coefficients in quantizing fields. A specific example was
given in our earlier work~\cite{Huang:2009ue}.

It is instructive to compare our results to those for an isotropic, unmagnetized ($B=0$)
fluid. In the isotropic case, the most general tensor
decompositions for the viscosity tensor and thermal conductivity tensor
are, respectively,
\begin{eqnarray}
\label{isotensor} \w^{\m\n\a\b}&=&\w\lb
\D^{\m\a}\D^{\n\b}+\D^{\m\b}\D^{\n\a}-\frac{2}{3}\D^{\m\n}\D^{\a\b}\rb+\z\D^{\m\n}\D^{\a\b},\non
\k^{\m\n}&=&\k\D^{\m\n}.
\end{eqnarray}
The corresponding Kubo formulas are
\begin{eqnarray}
\label{isotensor} \z&=&-\frac{\pt}{\pt\o}\im
G_{\tilde{P}\tilde{P}}^R({\bf 0},\o)\Big|_{\o\ra0},\non
\w&=&-\frac{1}{10}\frac{\pt}{\pt\o}\im
G^R_{\hat{\t}^{\m\n},\hat{\t}_{\m\n}}({\bf 0},\o)\Big|_{\o\ra0},\non
\k&=&\frac{1}{3}\frac{\pt}{\pt\o}\im G_{\gh^\m,\gh_\m}^R({\bf
0},\o)\Big|_{\o\ra0},
\end{eqnarray}
or when written in the rest frame,
\begin{eqnarray}
\label{isotensor2} \z&=&-\frac{\pt}{\pt\o}\im
G_{\tilde{P}\tilde{P}}^R({\bf 0},\o)\Big|_{\o\ra0},\non
\w&=&-\frac{\pt}{\pt\o}\im G^R_{\th^{12},\th^{12}}({\bf
0},\o)\Big|_{\o\ra0},\non \k&=&-\frac{\pt}{\pt\o}\im
G_{\gh^1,\gh^1}^R({\bf 0},\o)\Big|_{\o\ra0},
\end{eqnarray}
where
\begin{eqnarray}
\hat{\t}^{\m\n}&\equiv&\lb\D_{\m\r}\D_{\n\s}-\frac{1}{3}\D_{\m\n}\D_{\r\s}\rb\th^{\r\s},\non
\tilde{P}&\equiv&\hat{P}-\lb\frac{\pt P}{\pt \ve}\rb_n
\hat{\ve}-\lb\frac{\pt P}{\pt n}\rb_\ve \hat{n},\non
\hat{P}&\equiv&\frac{1}{3}\sum_i\th^{ii}.
\end{eqnarray}
These expressions are obtained directly from
\eq{kubovis4} and \eq{restkubo} by taking the limit $B=0$. However,
it is to be noted that even in the isotropic case one has
$(\tilde{P}_\perp,\tilde{P}_\parallel)\neq (\tilde{P},\tilde{P})$
for pressure-pressure correlation functions.

Thus, we have succeeded in expressing the transport coefficients of
a strongly magnetized fluid in terms of certain correlation functions
of quantities appearing in the energy-momentum tensor. For practical
evaluation of these correlation functions, one needs to specify a suitable
Lagrangian describing the system and a many-body scheme which allows for
resummation of polarization-type diagrams in a controlled manner.
Typically, non-perturbative resummation schemes are needed to obtain
correct infrared behavior of the transport coefficients.
Concrete applications have been presented, for example, in
Refs.~\cite{Jeon:1994if,FernandezFraile:2008vu}.

\section {Summary and Discussion}\label{summary}         %
In this work we have extended the development of covariant
magnetohydrodynamics of relativistic fluids initiated in
Ref.~\cite{Huang:2009ue}.  A hydrodynamical description of
physical processes in highly magnetized compact objects can be
achieved starting from conservation laws for conserved charges
and a suitable expansion scheme that systematically resolves the
large-length-scale and low-frequency limits.
In the most general case, the dissipative
function contains five shear viscosities,
two bulk viscosities, and three thermal conductivity coefficients,
which encode the transport processes at the microscopic level.
By utilizing Zubarev's  method of non-equilibrium statistical
operators, we have related these transport coefficients to correlation
functions of quantities entering the energy-momentum tensor in
equilibrium.  This is done to linear order in the expansion of
non-equilibrium statistical operators with respect to the gradients
of relevant statistical parameters (temperature, chemical potentials,
and momentum.) The linear perturbation theory recovers the ordinary
dissipative Navier-Stokes theory. In general, in order to obtain stable and causal
hydrodynamics one needs to expand to second order with respect to
the gradients, in which case additional transport coefficients
arise. It would be an interesting task to compute these coefficients in
the magnetohydrodynamic theory in the limit of large magnetic fields.

We have shown how to relate the correlation function, constructed from
ingredients of the energy-momentum tensor, to the retarded Green's
functions of equilibrium theory at vanishing momentum. The static
coefficients are obtained, as usual, by taking the limit of vanishing
frequency. In order to obtain an explicit form for the transport coefficients,
the energy-momentum tensor (or, equivalently, the Lagrangian)
describing the system needs to be specified. The two-particle response
functions computed at any finite order in the perturbation theory do not
give the correct dependence of the transport coefficients on the
coupling(s) of the underlying theory.  Therefore, a specific
resummation scheme is needed in each case.  The Kubo formulas
obtained in this work provide the necessary starting point for
implementing such a program for various systems affected by strong
magnetic fields.

\section*{Acknowledgments} We are grateful to T.~Brauner, D.~Fernandez-Fraile, Mei
Huang, and T.~Koide for helpful discussions and to John W.~Clark for
reading the manuscript and comments. This work was supported in part
by the Helmholtz Alliance Program of the Helmholtz Association,
contract HA216/EMMI ``Extremes of Density and Temperature: Cosmic
Matter in the Laboratory'', the Helmholtz International Center for
FAIR within the framework of the LOEWE
program launched by the State of Hesse, by the Deutsche Forschugsgemeinschaft
(Grant SE 1836/1-2) and CompStar, a research networking
programme of the European Science Foundation.

\appendix

\section{Dependence of $b^{\m\a}b^{\n\b}+b^{\m\b}b^{\n\a}$ on
  combinations in \eq{combination}}
\label{eight}
In a previous paper~\cite{Huang:2009ue}, we have treated the
combination $b^{\m\a}b^{\n\b}+b^{\m\b}b^{\n\a}$ as an independent tensor
structure, since it has the symmetries of the viscous tensor $\w^{\m\n\a\b}$.
Here we  show that, in fact, it can be written as a linear combination
of the tensors appearing in \eq{combination}. By direct calculation we
find
\begin{eqnarray}
b_{\m\a}b_{\n\b}&=&b^\l u^\x b^\r
u^\s\e_{\m\a\l\x}\e_{\n\b\r\s}=-b^\l u^\x b^\r
u^\s\lv\;
\begin{array}{cccc} g_{\m\n }  & g_{\m\b}  & g_{\m\r} & g_{\a\s}\\
                          g_{\a\n}   & g_{\a\b }  & g_{\a\r}  & g_{\a\s}\\
                          g_{\l\n}    & g_{\l\b}    & g_{\l\r}   & g_{\l\s}\\
                          g_{\x\n}   & g_{\x\b }   & g_{\x\r}   & g_{\x\s}\\
\end{array}\;\rv\non
&=&g_{\m\n}g_{\a\b}-g_{\m\b}g_{\n\a}+b_\b\lb b_\a g_{\m\n}-b_\m
g_{\n\a}\rb-u_\b\lb u_\a g_{\m\n}-u_\m g_{\n\a}\rb
\non&&-b_\n\lb b_\a
g_{\m\b}-b_\m g_{\a\b}\rb+u_\n\lb u_\a g_{\m\b}-u_\m
g_{\a\b}\rb\non&&-\lb b_\m u_\a-b_\a u_\m\rb\lb b_\n u_\b-b_\b
u_\n\rb\non&=&\D_{\m\n}\D_{\a\b}-\D_{\n\a}\D_{\m\b}+b_\a
b_\b\D_{\m\n}+b_\m b_\n\D_{\a\b}-b_\m b_\b\D_{\n\a}-b_\n
b_\a\D_{\m\b}.\non
\end{eqnarray}
Hence we have
\begin{eqnarray}
b_{\m\a}b_{\n\b}+b_{\n\a}b_{\m\b}&=&2\D_{\m\n}\D_{\a\b}-\D_{\n\a}\D_{\m\b}-\D_{\m\a}\D_{\n\b}+2b_\a
b_\b\D_{\m\n}+2b_\m b_\n\D_{\a\b}\non&&-b_\m b_\b\D_{\n\a}-b_\n
b_\a\D_{\m\b}-b_\n b_\b\D_{\m\a}-b_\m b_\a\D_{\n\b}\non&=&2 ({\rm
i}) - ({\rm ii})+2 ({\rm iii})-({\rm v}).
\end{eqnarray}

\section{Derivation of \eq{c}} \label{derivation}
The purpose of this Appendix is to give the details of the
derivation of  \eq{c}. We first establish the following lemmas. \\
{\bf Lemma 1}: The derivatives of $\b=1/T$ and $\a=\b \m$ in
thermal equilibrium can be written as
\begin{eqnarray}
\label{lemma1a}
D \b&=&\b\lb\H_\b\h+\F_\b\f\rb,\non
\label{lemma1b}
D\a&=&-\b\lb\H_\a\h+\F_\a\f\rb,
\end{eqnarray}
where
\begin{eqnarray}
\H_\b&\equiv&\lb\frac{\pt P}{\pt\ve}\rb_{n,B},\;\;\;
\F_\b\equiv-B\lb\frac{\pt M}{\pt
\ve}\rb_{n,B},\non \H_\a&\equiv&\lb\frac{\pt
P}{\pt n}\rb_{\ve,B},\;\;\; \F_\a\equiv-B\lb\frac{\pt M}{\pt
n}\rb_{\ve,B}.
\end{eqnarray}
{\bf Proof}: First, using $\b d\m=d\a-\m d\b$, $d\b=-\b^2 dT$, and $\ve+P=Ts+\m n$ we have
\begin{eqnarray}
dP&=&sdT+nd\m+MdB\non&=&-T(\ve+P)d\b+Tnd\a+MdB,
\end{eqnarray}
or, equivalently,
\begin{eqnarray}
\lb\frac{\pt P}{\pt\b}\rb_{\a,B}=-T(\ve+P),\;\;\;\;\;\; \lb\frac{\pt P}{\pt\a}\rb_{\b,B}=Tn,\;\;\;\;\;\lb\frac{\pt P}{\pt B}\rb_{\a,\b}=M.
\end{eqnarray}
Secondly, using $ds=\b d\ve-\a dn+\b MdB$, we obtain
\begin{eqnarray}
&&\lb\frac{\pt \b}{\pt n}\rb_{\ve,B}=-\lb\frac{\pt
  \a}{\pt\ve}\rb_{n,B},\;\lb\frac{\pt \b}{\pt
  B}\rb_{\ve,n}=\lb\frac{\pt(\b M)}{\pt\ve}\rb_{n,B},\;\non
&& \lb\frac{\pt \a}{\pt B}\rb_{\ve,n}=-\lb\frac{\pt(\b M)}{\pt n}\rb_{\ve,B}.
\end{eqnarray}
Based on the above thermodynamic relations, we may now write
\begin{eqnarray}
D\b&=&\lb\frac{\pt\b}{\pt\ve}\rb_{n,B}D\ve+\lb\frac{\pt\b}{\pt
n}\rb_{\ve,B}Dn+\lb\frac{\pt\b}{\pt
B}\rb_{\ve,n}DB\non&=&-\lb\frac{\pt\b}{\pt\ve}\rb_{n,B}(\ve+P)\h-\lb\frac{\pt\b}{\pt
n}\rb_{\ve,B}n\h\non&&-\lb\frac{\pt\b}{\pt\ve}\rb_{n,B}MDB+\lb\frac{\pt\b}{\pt
B}\rb_{\ve,n}DB\non&=&\lb\frac{\pt\b}{\pt\ve}\frac{\pt
P}{\pt\b}-\frac{\pt\b}{\pt n}\frac{\pt
P}{\pt\a}\rb_B\b\h\non&&
-\lb\frac{\pt\b}{\pt\ve}\rb_{n,B}MDB+\lb\frac{\pt\b}{\pt
B}\rb_{\ve,n}DB\non&=&\b\lb\frac{\pt
P}{\pt\ve}\rb_{n,B}\h+\b\lb\frac{\pt M}{\pt
\ve}\rb_{n,B}DB,
\end{eqnarray}
\begin{eqnarray}
D\a&=&\lb\frac{\pt\a}{\pt\ve}\rb_{n,B}D\ve+\lb\frac{\pt\a}{\pt
n}\rb_{\ve,B}Dn+\lb\frac{\pt\a}{\pt
B}\rb_{\ve,n}DB\non&=&-\lb\frac{\pt\a}{\pt\ve}\rb_{n,B}(\ve+P)\h-\lb\frac{\pt\a}{\pt
n}\rb_{\ve,B}n\h\non&&-\lb\frac{\pt\a}{\pt\ve}\rb_{n,B}MDB+\lb\frac{\pt\a}{\pt
B}\rb_{\ve,n}DB\non&=&\lb\frac{\pt\a}{\pt\ve}\frac{\pt
P}{\pt\b}-\frac{\pt\a}{\pt n}\frac{\pt
P}{\pt\a}\rb_B\b\h-\lb\frac{\pt\a}{\pt\ve}\rb_{n,B}MDB+\lb\frac{\pt\a}{\pt
B}\rb_{\ve,n}DB\non&=&-\b\lb\frac{\pt
P}{\pt n}\rb_{\ve,B}\h-\b\lb\frac{\pt M}{\pt
n}\rb_{\ve,B}DB.
\end{eqnarray}
From these equations,
we immediately recover Eqs.~(\ref{lemma1a}).
\\{\bf Lemma 2}: If the electric field is neglected, we have
\begin{eqnarray}
\label{alp1}
b^\n\pt_\n\a&=&\frac{\ve+P}{n T^2}\lb Tb_\n Du^\n-b^\n\pt_\n T\rb,\\
\label{alp2}
\Xi^{\r\n}\pt_\n\a&=&\frac{\ve+P}{n T^2}\lb T\Xi^{\r\n}
Du_\n-\Xi^{\r\n}\pt_\n T\rb,
\end{eqnarray}
or equivalently,
\begin{eqnarray}
\nabla_\m\a&=&\frac{\ve+P}{n T^2}\lb T
Du_\m-\nabla_\m T\rb.
\end{eqnarray}
{\bf Proof}: When the electric field is neglected, $E^\m=0$, we have
$\pt_\m T_{\rm EM}^{\m\n}=-n_\m F^{\m\n}=nE^\n=0$. Substituting
$T_{\rm EM}^{\m\n}=\frac{1}{2}B^2(u^\m u^\n-\Xi^{\m\n}-b^\m b^\n)$
for the energy-momentum tensor of the electromagnetic field,
a direct calculation shows that
\begin{eqnarray}
\label{elecmag}
\Xi_{\m\n}\lb\pt^\n \ln B-D u^\n+b^\r\pt_\r b^\n\rb=0.
\end{eqnarray}
From the first Maxwell equation, $\pt_\n\lb B^\m u^\n-B^\n u^\m\rb=0$, we have
\begin{eqnarray}
 u_\m\pt_\n\lb B^\m u^\n-B^\n u^\m\rb=0=u^\m u^\n \pt_\n B_\m-\pt_\n
 B^\n,
\end{eqnarray}
which implies
\begin{eqnarray}
\label{axi1} \pt_\n b^\n=-b^\n\pt_\n\ln B-b^\m D u_\m.
\end{eqnarray}
Contracting the second Maxwell equation, $\pt_\m H^{\m\n}=n^\n$, with
$b^{\r\n}$, we obtain
\begin{eqnarray}
0&=&b^{\r\n}b_{\m\n}\pt^\m H+H b^{\r\n}\pt^\m b_{\m\n}\non
&=&\Xi^{\r\m}\pt_\m H+H \e^{\r\n\l\s}\e_{\m\n\a\b}b_\l
u_\s(u^\b\pt^\m b^\a +b^\a\pt^\m u^\b)\non&=&\Xi^{\r\m}\pt_\m H+H\lb
b^\m\pt_\m b^\r+u^\r b^\s b^\m\pt_\m u_\s-Du^\r-b^\r b^\l
Du_\l\rb\non&=&\Xi^{\r\m}\pt_\m H+H\Xi^{\r\m}\lb b^\n\pt_\n
b_\m-Du_\m\rb,
\end{eqnarray}
which implies
\begin{eqnarray}
\label{axi2} \Xi_{\n\r}b^\m\pt_\m
b^\r=\Xi_{\n\r}Du^\r-\Xi_{\n\m}\pt^\m \ln H.
\end{eqnarray}
Substituting \eq{elecmag} into \eq{axi2}, we obtain
\begin{eqnarray}
\label{axi3} \Xi_{\n\r}b^\m\pt_\m
b^\r=\Xi_{\n\r}\lb Du^\r-\pt^\r \ln M\rb.
\end{eqnarray}
The equation of motion of an ideal fluid is
\begin{eqnarray}
0&=&\pt_\m T^{\m\n}_0=\pt_\m T^{\m\n}_{\rm F0}\non&=& u^\n
D\ve+(\ve+\pperp) u^\n\h+(\ve+\pperp) D
u^\n\non&&
-\Xi^{\n\m}\pt_\m\pperp+MB(b^\n\pt_\m b^\m+b^\m\pt_\m b^\n)+b^\m
b^\n\pt_\m\ppara.\non
\end{eqnarray}
Contracting this equation with $b_\n$ and $\Xi_{\r\n}$ yields
\begin{eqnarray}
\label{aeom1} &(\ve+\pperp) b_\n D u^\n-MB\pt_\m b^\m-b^\m
\pt_\m\ppara=0,&\\
\label{aeom2} &(\ve+\pperp)\Xi_{\r\n} D
u^\n-\Xi_{\r\m}\pt^\m\pperp+MB\Xi_{\r\n}b^\m\pt_\m b^\n=0.&
\end{eqnarray}
Substitution of \eq{axi1} into \eq{aeom1} followed by some manipulation
gives
\begin{eqnarray}
(\ve+\pperp) b_\n D u^\n+MB(b^\n\pt_\n\ln B+b^\n Du_\n)-b^\m
(s\pt_\m T+n\pt_\m \m+M\pt_\m B)=0.\nonumber
\end{eqnarray}
We then have
\begin{eqnarray}
nb^\n\pt_\n\m=(\m n+Ts)b^\n Du_\n-sb^\n\pt_\n T,
\end{eqnarray}
or equivalently,
\begin{eqnarray}
b^\n\pt_\n\a=\frac{\ve+P}{n T^2}\lb Tb_\n Du^\n-b^\n\pt_\n T\rb,
\end{eqnarray}
thereupon proving \eq{alp1}.
Substituting \eq{axi3} into \eq{aeom2}, we obtain
\begin{eqnarray}
(\ve+P) \Xi_{\r\n}D u^\n&=&\Xi_{\r\n}\lb s\pt^\n T+n\pt^\n
\m\rb,
\end{eqnarray}
or equivalently,
\begin{eqnarray}
\Xi^{\r\n}\pt_\n\a=\frac{\ve+P}{n T^2}\lb T\Xi^{\r\n}
Du_\n-\Xi^{\r\n}\pt_\n T\rb,
\end{eqnarray}
which proves \eq{alp2}.

Making use of Lemmas 1 and 2, we may now write
\begin{eqnarray}
\pt_\m\b_\n&=&\b\pt_\m u_\n+u_\n\pt_\m\b\non&=&\b u_\m\D_{\n\r}
Du^\r+\b\nabla_\m u_\n+u_\m u_\n
D\b+u_\n\D_{\m\r}\pt^\r\b\non&=&\b^2\lb Tu_\m\D_{\n\r}
Du^\r-u_\n\D_{\m\r}\pt^\r T\rb\non&+&\b\lb\D_{\m\r}\D_{\n\s}+\H_\b u_\m
u_\n\D_{\r\s}+\F_\b u_\m u_\n\Xi_{\r\s}\rb\pt^\r u^\s,
\end{eqnarray}
\begin{eqnarray}
\pt_\m\a&=&\D_{\m\n}\pt^\n\a+u_\m D\a=\frac{\ve+P}{n T^2}\lb
T\D_{\m\n} Du^\n-\D^{\m\n}\pt^\n T\rb\non&&-\b
u_\m\lb\H_\a\D_{\r\s}+\F_\a\Xi_{\r\s}\rb\pt^\r u^\s.
\end{eqnarray}
Substituting these relations into
\begin{eqnarray}
\ch&\equiv&\th_{\m\n}(\bx,t)\pt^\m
\b^\n(\bx,t)-\nh^\m(\bx,t)\pt_\m\a(\bx,t),
\end{eqnarray}
we immediately obtain
\begin{eqnarray}
\ch&=&\b^2\lb u_\m
\th^{\m\n}\D_{\n\r}-\frac{\ve+P}{n}\nh^\n\D_{\n\r}\rb(TDu^\r-\nabla^\r
T)\non&+&\b\Bigg[\th^{\m\n}\lb\D_{\m\r}\D_{\n\s}+\H_\b u_\m
u_\n\D_{\r\s}+\F_\b u_\m u_\n\Xi_{\r\s}\rb\non&&+\nh^\m
u_\m\lb\H_\a\D_{\r\s}+\F_\a\Xi_{\r\s}\rb\Bigg]\nabla^\r u^\s\non
&=&\lb\frac{n}{\ve+P}u_\m
\th^{\m\n}\D_{\n\r}-\nh^\n\D_{\n\r}\rb\nabla^\r\a\non&+&
\b\Bigg[\th^{\m\n}\lb\D_{\m\r}\D_{\n\s}+\H_\b
u_\m u_\n\D_{\r\s}+\F_\b u_\m u_\n\Xi_{\r\s}\rb\non&&+\nh^\m
u_\m\lb\H_\a\D_{\r\s}+\F_\a\Xi_{\r\s}\rb\Bigg] w^{\r\s},\non
\end{eqnarray}
with $w^{\r\s}\equiv(\nabla^\r u^\s+\nabla^\s u^\r)/2$.

\section{Integration of Kubo Correlator and Retarded Green's Function} \label{kubogreen}
In this Appendix, we derive the relation between the space-time integral
of the Kubo correlator and the retarded Green's function. This relation was
used to derive \eqs{kubovis2}{kuboheat2}. Let us denote the integral
of a general Kubo correlator by
\begin{eqnarray}
I\equiv\int d^3\bx'\int_{-\infty}^t dt' e^{\ve(t'-t)}\lb{\hat
X}(\bx,t),{\hat Y}(\bx',t')\rb.
\end{eqnarray}
First, we rewrite the Kubo correlator in the following form, assuming
that $\lim_{t'\ra-\infty}\lan{\hat X}(\bx,t){\hat Y}(\bx',t')\ran_l
=\lan{\hat X}(\bx,t)\ran_l\lan{\hat Y}(\bx',t')\ran_l$ so that
the correlation vanishes at $t'\ra-\infty$:
\begin{eqnarray}
\lb{\hat X}(\bx,t),{\hat Y}(\bx',t')\rb&\equiv&\int_0^1 d\t\lan
{\hat X}(\bx,t)\ls e^{-\ah\t}{\hat Y}(\bx',t')e^{\ah\t}-\lan{\hat
Y}(\bx',t')\ran_l\rs\ran_l\non &=&\frac{1}{\b}\int_0^\b d\t\lan
{\hat X}(\bx,t)\ls {\hat Y}(\bx',t'+i\t)-\lan{\hat
Y}(\bx',t'+i\t)\ran_l\rs\ran_l\non&=&\frac{1}{\b}\int_0^\b
d\t\int_{-\infty}^{t'}ds
\non&&\frac{d}{ds}\lan {\hat X}(\bx,t)\ls {\hat Y}(\bx',s+i\t)-\lan{\hat
Y}(\bx',s+i\t)\ran_l\rs\ran_l\non&=&\frac{i}{\b}\int_{-\infty}^{t'}ds\lan
\ls{\hat X}(\bx,t), {\hat Y}(\bx',s)\rs\ran_l.
\end{eqnarray}
In order to obtain the last equality we have used the
Kubo-Martin-Schwinger relation $\lan{\hat X}(t){\hat
Y}(t'+i\b)\ran_l=\lan{\hat Y}(t'){\hat X}(t)\ran_l$.

Then we find that
\begin{eqnarray}
I&=&\int d^3\bx'\int_{-\infty}^t dt'
e^{\ve(t'-t)}\frac{i}{\b}\int_{-\infty}^{t'}ds\lan \ls{\hat
X}(\bx,t), {\hat Y}(\bx',s)\rs\ran_l\non&=&-\int
d^3\bx'\int_{-\infty}^t dt'
e^{\ve(t'-t)}\frac{1}{\b}\int_{-\infty}^{t'}ds
G_R(\bx-\bx',t-s)\non&=&-\int_{-\infty}^0 dt' e^{\ve
t'}\frac{1}{\b}\int_{-\infty}^{t'}ds\int\frac{d\o}{2\p} e^{i\o
s}\lim_{\bk\ra0}G_R(\bk,\o)\non&=&-\frac{1}{\b}\int\frac{d\o}{2\p}\frac{1}{i\o(i\o+\ve)}
\lim_{\bk\ra0}G_R(\bk,\o)\non&=&
-\frac{1}{\b}\oint\frac{d\o}{2\p}\frac{1}{i\o(i\o+\ve)}
\lim_{\o\ra0}\lim_{\bk\ra0}\ls
G_R(\bk,0)+ \o\frac{\pt}{\pt\o}G_R(\bk,\o)
+\cdots\rs\non
&=&\frac{i}{\b} \lim_{\o\ra0}
\lim_{\bk\ra0}\frac{\pt}{\pt\o}G_R(\bk,\o),
\end{eqnarray}
the contour integration being performed in the upper-half plane
where the retarded Green's function is analytic.
If $\ah=\bh$, the real (imaginary) part of the retarded Green's
function is an even (odd) function of $\o$,
\begin{eqnarray}
I=-\frac{1}{\b} \lim_{\o\ra0}\lim_{\bk\ra0}\frac{\pt}{\pt\o}\im
G_R(\bk,\o).
\end{eqnarray}
This result is used in Sec.~\ref{nonequilibrium} to obtain
\eqs{kubovis2}{kuboheat2}.

\bibliographystyle{model1-num-names}

\end{document}